\UseRawInputEncoding
\documentclass[floats,amsmath,amssymb,prl,aps,twocolumn]{revtex4}

\usepackage{graphics,amsmath,graphicx}
\usepackage{txfonts}
\usepackage[colorlinks=true,citecolor=blue,linkcolor=blue,urlcolor=blue]{hyperref}

\usepackage{dcolumn} 
\usepackage{bm} 
\usepackage{physics}
\usepackage{tikz}
\usepackage{subfigure}
\usepackage{babel}

\begin{document}
\usetikzlibrary{calc,backgrounds,arrows.meta}

\pgfdeclarelayer{background}
\pgfdeclarelayer{foreground}
\pgfsetlayers{background,main,foreground}


\title{Pseudogap metal induced by long-range Coulomb interactions}

\author{K. Driscoll, A. Ralko, S. Fratini$^*$} 
\affiliation{Institut N\'eel, CNRS \& Univ. Grenoble Alpes, 38042 Grenoble, France}

\begin{abstract}
In correlated electron systems the metallic character of a material can be strongly suppressed near an integer concentration of conduction electrons as Coulomb interactions forbid the double occupancy of local atomic orbitals.
While the Mott-Hubbard physics arising from such on-site interactions has been largely studied, several unexplained phenomena observed in correlated materials challenge this description and call for the development of new ideas. Here we explore a general route for obtaining correlated behavior that is decidedly different from the spin-related Mott-Hubbard mechanism and instead relies on the presence of unscreened,  long-range Coulomb interactions. We find a previously unreported pseudogap metal phase characterized by a divergent quasiparticle mass and the opening of a Coulomb pseudogap in the electronic spectrum. The destruction of the Fermi liquid state occurs because the electrons move 
in a nearly frozen, disordered charge background,  as collective charge rearrangements are drastically slowed down by the frustrating nature of long-range potentials on discrete lattices.
The present pseudogap metal realizes an early conjecture by Efros, that a soft Coulomb gap should appear for quantum lattice electrons with strong unscreened interactions due to self-generated randomness.
\end{abstract}

\date{\today}
\maketitle


The Mott metal-insulator transition (MIT) is one of the cornerstones of modern condensed matter physics 
\cite{Imada98,Georges96}. Originally devised to explain 
the cause of the insulating state in narrow-band materials with partially-filled bands,
modern focus has 
shifted to understanding 
the anomalous properties of metals that arise near the MIT and their possible consequences in stabilizing other phases, including superconductivity.
Most theoretical developments in the field have relied on the Hubbard model and its variants, where the Coulomb repulsion between electrons is reduced to its strongest (on-site) term --- neglecting all non-local terms from the outset.
Based on these models and the techniques that have been developed and applied to the problem,
we now have a broad understanding of
the physics of strongly correlated electron systems.
Despite this widespread success several experimental puzzles in quantum materials remain however unexplained \cite{Keimer15,Zheng17,Bruin13,Delacretaz17,Padhi18}, which drives us to revisit the implicit assumptions in the Mott-Hubbard description.
To this aim we solve a lattice model which explicitly includes long-range electron-electron interactions, demonstrating that these can give rise to strongly correlated behavior physically distinct from the Mott type. 
Unrelated to the spin degrees of freedom, we find strong mass renormalization caused by the buildup of  
non-local charge correlations dressing the quasiparticles, which is 
accompanied by the opening of a pseudogap in the single-particle  spectrum. This happens at the approach of Wigner crystallization,
where the fluctuating charge density is collectively slowed down and behaves effectively as a nearly frozen random medium, thereby enabling the Efros-Shklovskii Coulomb gap phenomenon.

\paragraph{Model and methods.}

We study spinless electrons interacting through a long-range repulsive potential $V(R)=V \times (R/a)^{-\alpha}$ on a two-dimensional lattice, as described by the
following Hamiltonian \cite{Pramudya11}:
\begin{equation}
H=- t \sum_{\langle i j \rangle}  c_{i}^{\dagger} c_{j}+\frac{1}{2} 
\sum_{i j} V(R_{i j})(\hat{n}_{i} - n)(\hat{n}_{j}- n).
\label{eq:H}
\end{equation}
Here $c_{i}^{\dagger}$ and $c_i$ are  creation and annihilation operators for electrons on local atomic orbitals, $\hat{n}_{i}$ is the local density operator, $t$ is the hopping matrix element between nearest neighbor sites, which we take to be isotropic, and $n=1/2$ is the average electron concentration. The strength of the interactions is controlled by $V$, the value of the potential at one lattice spacing $a$ (which we set as the unit length). For illustrative purposes we choose to present results for the  triangular lattice, but our findings are not specific to this particular lattice geometry \cite{SM}.  
We explore the full phase diagram of the model, taking the power-law exponent $\alpha$ as a continuous parameter. The chosen form of $V(R)$ includes the pristine Coulomb potential $V(R)\sim 1/R$ ($\alpha=1$)
and the commonly studied nearest-neighbor repulsion characteristic of the extended Hubbard model ($\alpha=\infty$), as well as  the  dipolar form $V(R)\sim 1/R^3$  of the two-dimensional electron gas near a metallic gate  ($\alpha=3$). 

Eq. (\ref{eq:H}) is solved numerically via both Lanczos and brute force exact diagonalization at zero temperature on finite-size clusters with $N_s=12,18,24$ sites \cite{Mambrini}. Finite-size errors on the kinetic part are minimized by averaging over twisted boundary conditions (TBC) both at fixed particle number \cite{Poilblanc91}
 and in the grand-canonical ensemble \cite{Gros92,Koretsune}, which restores the exact $N_s\to \infty$ result in the non-interacting limit, $V/t\to 0$ (see SI). For the interaction part, we extend the cluster size to the thermodynamic limit by considering infinitely repeated simulation cells. We  perform the corresponding lattice sums using the Ewald summation method \cite{Pramudya11},
which ensures that the  electrostatic (Madelung) energy of periodic configurations is exactly recovered in the classical limit, $t/V\to 0$. 
Details on the calculation of observables can be found in the SI file.

\paragraph{Phase diagram.}

\begin{figure}
\includegraphics[width=8.6cm]{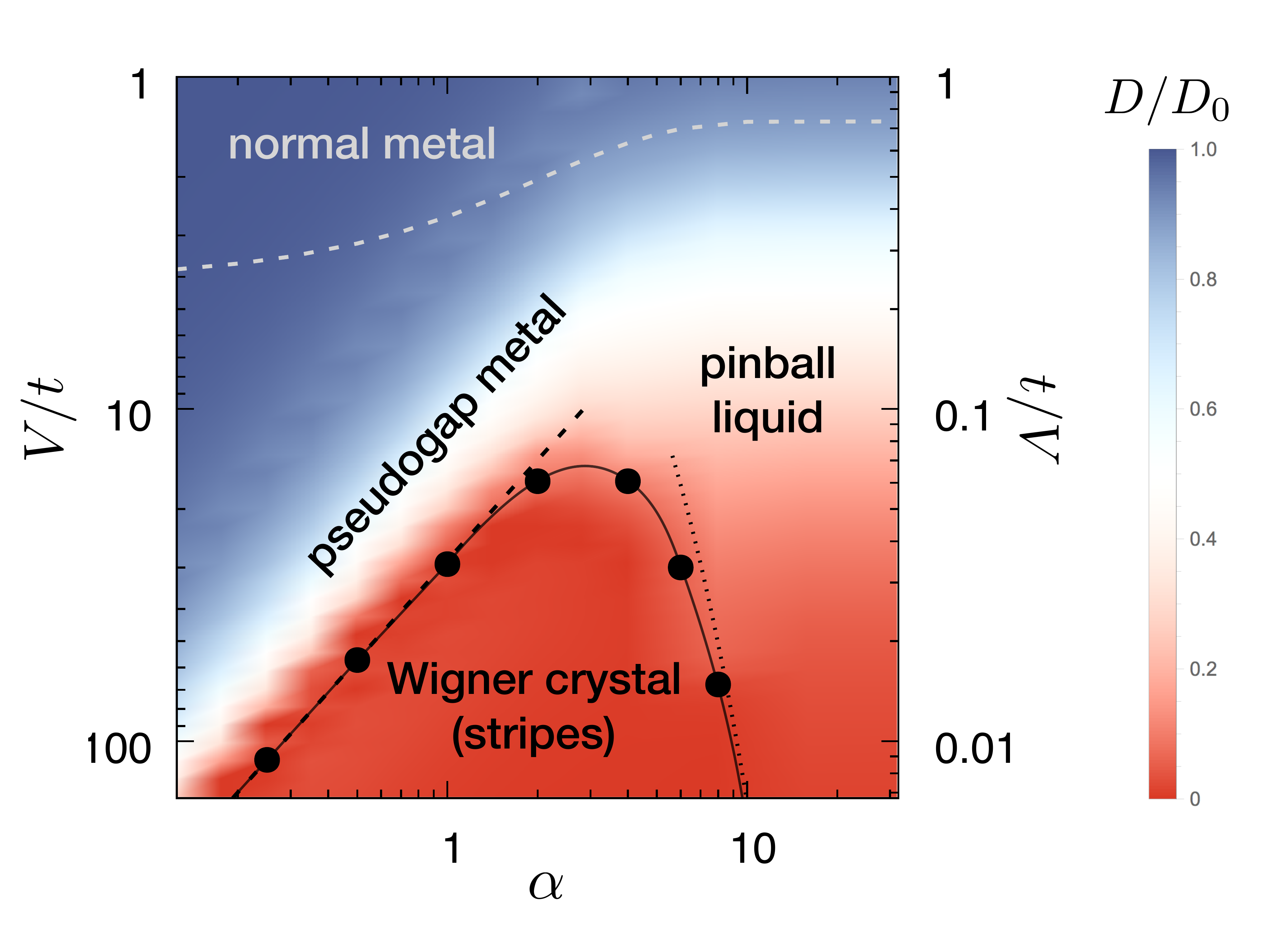}
\caption{\textbf{Phase diagram of the triangular lattice model with long-range interactions.}
 The line and the full symbols indicate the metal-insulator transition to a stripe-ordered Wigner crystal, signaled by  the vanishing of the Drude weight $D$ ($N_s=18$ sites, using $121$ twisted boundary conditions, TBC). The color map is the Drude weight (697 grid points, $N_s=12$ sites, $400$ TBC).
The gray dashed line is the charge ordering transition as obtained from the random phase approximation (RPA), which ignores correlations and does not capture the extreme fragility of the Wigner crystal. The black dashed and dotted lines are  strong-coupling estimates for the  Wigner crystal melting (see text).}
\label{fig:PD-ED}
\end{figure}

Fig. \ref{fig:PD-ED} presents the phase diagram of the model as a function of the power-law exponent $\alpha$. Four different regions are found: normal metal at weak interaction strengths, and the pinball liquid, stripe-ordered Wigner crystal and pseudogap metal at strong interactions. 
The origin of all three strongly interacting phases can be understood starting from the classical limit of the model. 
At $t/V=0$, for nearest-neighbor repulsive potentials ($\alpha\to \infty$, right side of Fig. \ref{fig:PD-ED}) 
there exist infinitely many classical configurations,  where part of the particles (``pins'') are located on a superlattice with threefold periodicity, the other particles (``balls'') being randomly distributed on the remaining honeycomb lattice \cite{Hotta06}, all having the same Madelung energy  $E_{Mad}/N_s=V/2$. This degeneracy is lifted by quantum fluctuations: as soon as $t/V>0$ ($V/t < \infty$), minimizing the kinetic term for the ``balls'' 
provides a net energy gain $\propto t$, identifying a unique macroscopic ground state --- the pinball liquid (PL) \cite{Hotta06}. 
This state has strong threefold correlations reminiscent of the classical limit and a weakly metallic character, which progressively evolves into a normal metal upon reduction of the interaction strength $V/t$ 
\cite{Miyazaki09}.

\begin{figure*}
\includegraphics[width=18cm]{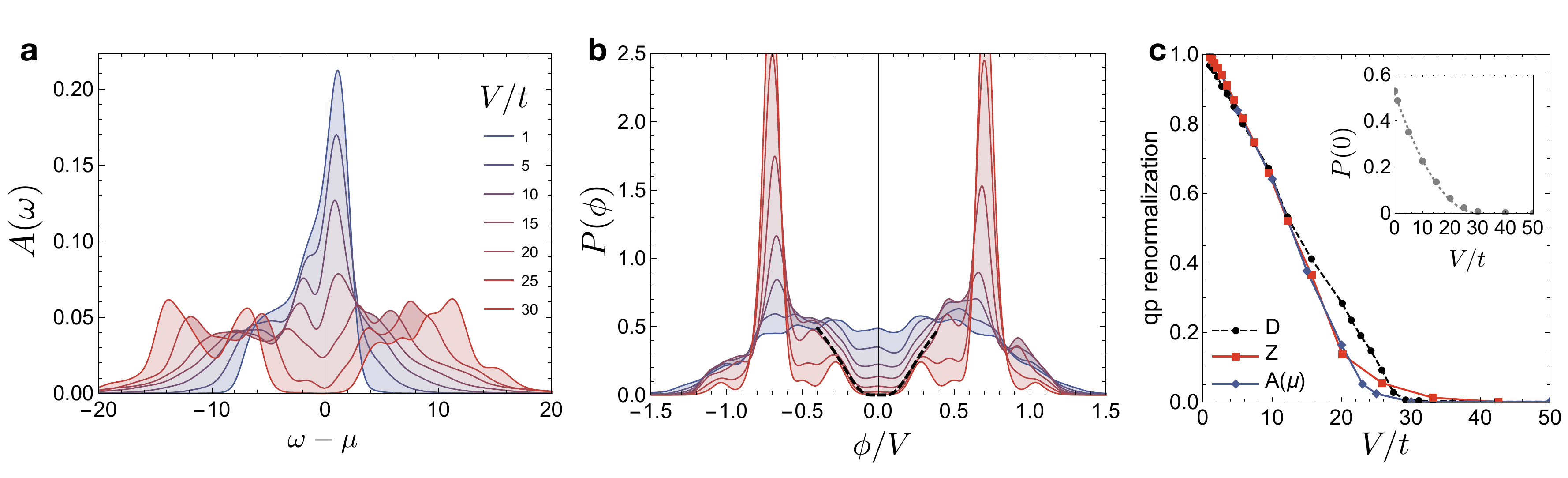}
\caption{\textbf{Pseudogap metal.} \textbf{a}, Spectral function $A(\omega)$ at $\alpha=1$ and $N_s=18$ and averaged over $16$ TBC, illustrating the pseudogap phenomenon. All spectra are smoothened by a Gaussian broadening $\delta=0.7t$. \textbf{b},  Distribution of classical local potentials $P(\phi)$ (broadening $\delta=0.05V$), showing the opening of the Coulomb gap; the dashed line is $P\sim e^{-V/\xi |\phi| }$ with $\xi=2$; the values of $V/t$ and the color code are the same as in (\textbf{a}).  \textbf{c}, 
Drude weight $D$  (averaged over $121$ TBC) normalized by the exact non-interacting value $D_0=0.247$,  quasiparticle weight $Z$ (angular-averaged over the Fermi surface) and DOS $A(\mu)$ at the Fermi energy, divided by the non-interacting value. The inset shows $P(0)$, tracking the plasma dip. The dashed line is $P(0)\sim |V-V_c|^2$. 
} 
\label{fig:pseudogap}
\end{figure*}

Long-range interactions also immediately lift the massive degeneracy characterizing the classical limit \cite{Mahmoudian15}. As soon as $\alpha < \infty$ (bottom part of Fig. \ref{fig:PD-ED}), the interactions beyond nearest neighbors favor linear stripe configurations, which become the most stable states at $t=0$. The stripe phase, which is the lattice analogue of a Wigner crystal \cite{Fratini09}, remains insulating in the presence of quantum fluctuations at small $t>0$, as indicated by the vanishing of the Drude weight (Fig. \ref{fig:PD-ED} and \ref{fig:pseudogap}(c)).  The pinball liquid can still be stabilized above some critical value of $t/V$ as long as $\alpha \gtrsim 2$. The potential energy difference  with the classically more stable stripe configurations behaves asymptotically as $\Delta E \propto V/(R_2)^\alpha $, with $R_2=\sqrt{3}$ the second neighbor distance. The transition from stripes to PL occurs when the kinetic energy gain associated with the itinerant carriers overcomes such  energy difference, leading to $(V/t)_c\propto 3^{\alpha/2}$ (dotted line in Fig. \ref{fig:PD-ED}).

\paragraph{Suppression of order by the long-range interactions.}
Reducing the long-range exponent below $\alpha \leq 2$  reveals a dome-like shape, with the stripe-ordered insulator becoming more and more unstable with increasing range of interactions.
The  fragility of Wigner crystal order is a  known  feature of long-range interactions in the continuum: in the jellium model with pure Coulomb repulsion ($\alpha=1$),
the ordered state melts due to the existence of extremely soft,
shear collective modes 
that are easily accessible via a  low energetic cost \cite{Andreev79,Spivak01,Pramudya11}.
For this reason,  the ratio of interaction to kinetic energy,  as given by  the appropriate dimensionless interaction parameter, is large at the  transition:  
$r_s\simeq 31$ for quantum electrons in $d=2$ \cite{DrummondPRL09}.
Strong interaction effects then naturally persist into the metallic state beyond melting, causing  short-range spatial correlations that are reminiscent of 
those in the ordered phase 
\cite{Mahan}, and a consequently large correlation energy. 

Analogously, for quantum 
lattice electrons as considered here, long-range interactions favor charge fluctuations, destabilizing the  Wigner crystal (stripe) order
and uncovering the correlated metallic state that lies underneath.  
To assess this effect, we again resort to the $t/V\to 0$ limit and evaluate the energy  required to create a defect of the ordered pattern, $E_{d}$ \cite{Fratini09}, which is obtained by displacing a carrier from its equilibrium position on the stripe to a neighboring unoccupied site on the lattice. While this energy cost is exactly $E_{d}= V$ in the nearest neighbor limit ($\alpha\to \infty$), 
it is steadily suppressed upon increasing the range of the interactions. As a result, defects are more and more easily created by quantum fluctuations when $t>0$.
The quantum melting transition  occurs through proliferation of such defects when $t \sim E_{d} $ \cite{Andreev69,Tsiper98,Fratini09}. 
From  the asymptotic expression $E_{d}\simeq  0.469 V \alpha$  we obtain $(V/t)_c \propto 1/\alpha$ for small $\alpha$, as observed in Fig. \ref{fig:PD-ED} (black dashed line). 
For comparison we show the transition predicted by the random phase approximation (gray dashed line) \cite{Cano11}. This approximation  captures the onset of local charge order but it completely misses the fragility of long-range order at small $\alpha$, which arises from correlations beyond mean-field level.
Remarkably, the pure Coulomb case [$\alpha=1$, $(V/t)_c \simeq 29$, corresponding to $r_s=7.2$] lies well on  the asymptotic ``small $\alpha$'' side of the dome: in this regime we have $E_d\ll V$, signifying that the local, short range energy scale $V$ and the global, long-range scale responsible for collective behavior and melting are indeed well separated. 
As we shall see,
this separation of energy scales has profound consequences on the electronic properties of the metal.

\paragraph{Pseudogap metal.}

Fig. \ref{fig:pseudogap}a illustrates the evolution as a function of $V/t$ of the local single-particle spectral function $A(\omega)$ in the metallic state. As the interaction strength increases, a pseudogap  opens at the Fermi energy ($\omega=\mu$), which progressively deepens and broadens as excitations move towards high energies, $\omega \sim V$.
 The density of states at the Fermi energy, $A(\mu)$,  falls approximately linearly with $V/t$, then flattens deep in the pseudogap phase and eventually vanishes at the MIT at $(V/t)_c \simeq 29$ (Fig. \ref{fig:pseudogap}c);  the  pseudogap coalesces into a hard gap in the stripe phase beyond this value.  Fig. S17shows analogous results obtained on the square lattice, demonstrating  that  the  source  of  frustration  responsible  for  the  pseudogap  formation originates from the long-range interactions,  and not the lattice geometry.

Concomitant with the development of the pseudogap in the one-particle spectrum, 
electronic correlations build up, signalled by a steady decrease of both the Drude weight, $D$, and the quasiparticle weight, $Z$, with the latter following closely the behavior of $A(\mu)$ (Fig. \ref{fig:pseudogap}c). At the MIT both $D$ and $Z$ vanish, indicating the divergence of the quasiparticle mass, $m^*/m_b\propto 1/Z$, and of the optical effective mass, $m^*_{opt}/m_b= D_0/D$ \cite{SM}.
Strikingly, the 
mechanism for mass divergence at work here is radically different 
from the spin-related mechanism involved in  the bandwidth-controlled Mott-Hubbard transition.
In the  case of the Mott-Hubbard MIT, the spectral function features a quasiparticle peak that remains pinned at the Fermi energy, and whose shrinking with $Z$ mostly causes the divergence of the effective mass \cite{Imada98,Georges96}.
Here no peak narrowing is found, and it is instead the value of the renormalized density of states (DOS) at the Fermi energy, $A(\mu)$,
that falls continuously to $0$ controlling the quasiparticle renormalization (Fig. \ref{fig:pseudogap}c).

\paragraph{Self-generated randomness and short range correlations.}

The pseudogap phenomenon revealed in the preceding paragraphs is strongly reminiscent of the soft Coulomb gap characteristic of disordered 
insulators. There, stability arguments  imply that the DOS of an interacting electron system in the presence of 
quenched disorder must vanish
at the Fermi energy \cite{Efros75}, 
due to their long-range mutual interactions. Similar physics was also reported in clean classical Coulomb liquids,
where it was shown that the long-distance potentials from electrons beyond the correlation length, when taken collectively, act as a source of (self-generated) randomness \cite{Efros92,Schmalian00,Pramudya11,Mahmoudian15,Rademaker18}. 
The observations presented in Fig. \ref{fig:pseudogap}  highlight  that the phenomenon of  self-generated randomness and the associated Coulomb pseudogap exist also in the clean \textit{quantum} case, as hypothesized by Efros almost three decades ago \cite{Efros92}.
The resemblance between the quantum phase diagram Fig. \ref{fig:PD-ED} and its classical analogue determined in Ref. \cite{Pramudya11} is striking.

To track the origin of the pseudogap, we determine the  distribution of  electrostatic  site energies in the quantum ground state $|\psi \rangle$, which can be evaluated as $P(\phi_i)=\langle \psi| \delta[\phi_i- \sum_{j\neq i} V(R_{ij}) (\hat{n}_j-n)]|\psi \rangle$ (the site index can be ignored as this quantity is translationally invariant in the present case). For classical electrons, $P(\phi)$ would reduce to the density of states studied in the Efros-Shklovskii soft gap problem \cite{Efros75}. 
In the quantum case, it represents the fluctuating 
background where the electron motion takes place. 
   
Fig. \ref{fig:pseudogap}b shows that, 
prior to the pseudogap opening observed in the full electronic spectrum, a broad dip develops already in the distribution of site potentials. Interestingly, its shape at the transition 
is compatible with that caused by short-range charge correlations in self-generated Coulomb glasses, $P\sim e^{-V/\xi |\phi| }$ (dashed line) \cite{Rademaker18}. There, the correlation hole that forms around  electrons in order to minimize their mutual interactions was shown to deplete the classical DOS below the Efros-Shklovskii (ES) bound, $P_{ES}\sim |\phi|$ ($P_{ES}\sim |\phi|^{d/\alpha-1}$ in the general case in $d$ dimensions and exponent $\alpha$). Such correlation hole, or ``electronic polaron'',  is a common feature of 
electron liquids with unscreened Coulomb interactions \cite{Mahan}. Its buildup  within the pseudogap phase is confirmed here by direct evaluation of the charge correlation function \cite{SM}.
We have verified that 
our method fully recovers the prediction $P_{ES}\sim |\phi|$ upon suppressing short-range correlations via the introduction of extrinsic (quenched) disorder, and that the pseudogap disappears for short range interactions $\alpha>D$, demonstrating that the observed  pseudogap is the consequence of the long-range Coloumb interaction.

\paragraph{Soft collective excitations.}

\begin{figure}
\includegraphics[width=8.5cm]{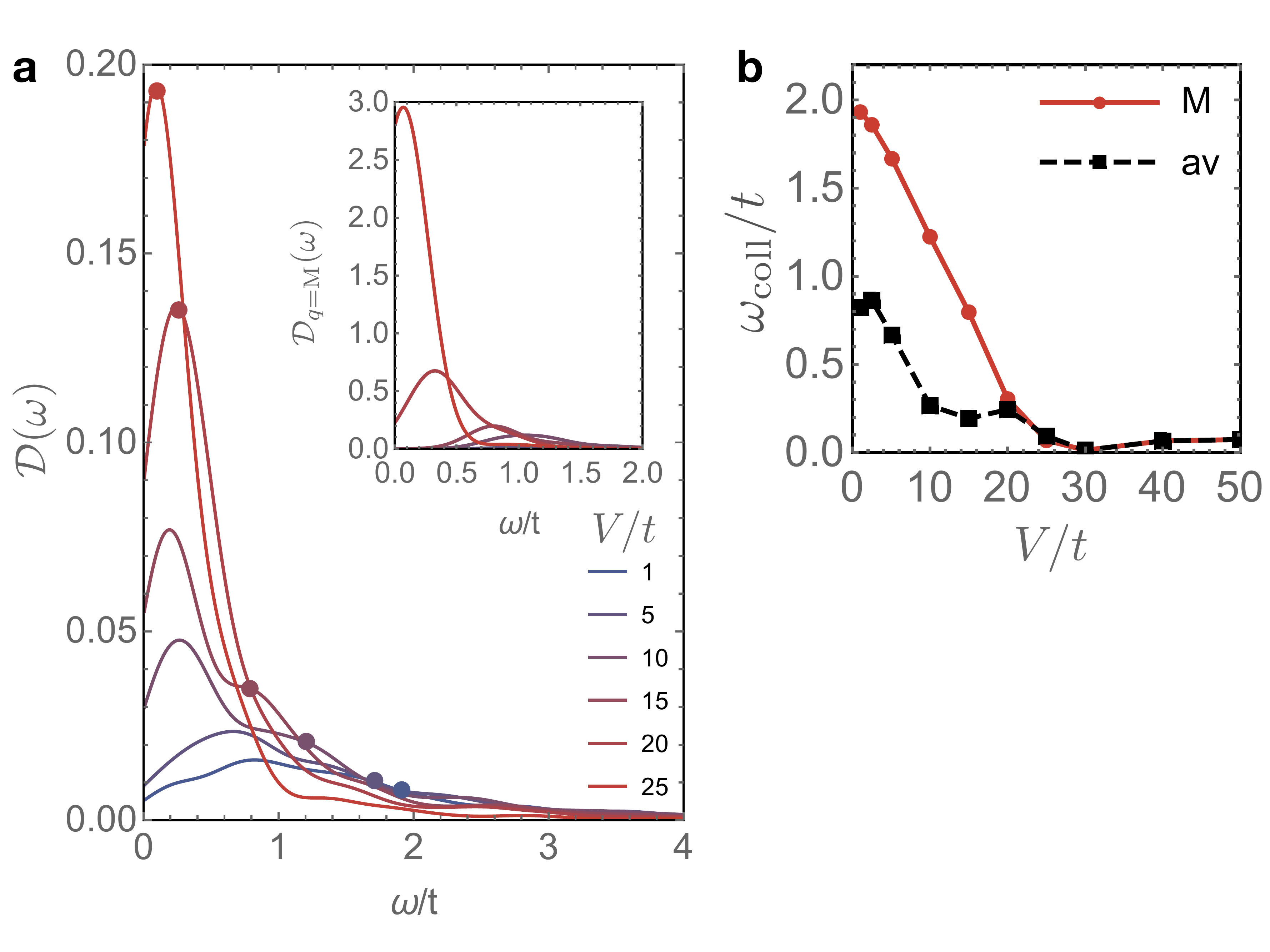}
\caption{\textbf{Soft collective excitations in the pseudogap phase.} \textbf{a} Spectral function $\mathcal{D}(\omega)$ of the charge fluctuations, averaged over the Brillouin zone (Gaussian broadening $\delta=0.2t$). The inset shows the same quantity evaluated at the stripe-ordering wavevector $q=M$; this critical mode is also visible as a shoulder in the  main panel (circles). \textbf{b} Frequency of the boson peak, controlling the timescale of the charge fluctuations.  
} 
\label{fig:boson}
\end{figure}

The results presented above demonstrate strongly correlated behavior arising from long-range charge-charge interactions, which is unrelated to the paradigmatic Mott mechanism. Our findings indicate that in systems with unscreened, long-range interactions, the collective charge fluctuations are able to provide a self-generated random environment, 
thereby enabling precursors of the Efros-Shklovskii Coulomb gap phenomenon. 
This fluctuating environment is polarizable and responds to the motion of the individual electrons, being ultimately responsible for the mass enhancement via the formation of electronic polarons.

The  existence of a Coulomb pseudogap 
necessarily implies that there 
is a marked separation of timescales between the (fast) motion of individual electrons and the (much slower) global rearrangements of the charge at long distances: the idea being that the charge fluctuation background is \textit{almost frozen}, being collectively jammed 
by the mutual interactions among its constituents \cite{Andreev79,Emery93,Schmalian00,Spivak01,Pramudya11}.
We can actually provide quantitative support to this statement, by evaluating the spectrum of charge fluctuations, $\mathcal{D}(\omega)=\sum_m|\langle m|\ \hat{\rho}_q |\psi\rangle|^2 \delta[\omega-(E_m-E_0)]$, where $\hat{\rho}_q$ is the Fourier transform of the charge density $\hat{n}_i$, and $m$ and $E_m$ are all the eigenstates and eigenenergies.
Fig. \ref{fig:boson}a shows that there is a strong contribution to the spectrum  that is soft throughout the pseudogap phase,  peaking at $\omega_{\mathrm{coll}}\simeq 0.2-0.25t$ (see also Fig. \ref{fig:boson}b). This is about $30-50$ times lower than the free-electron bandwidth, $9t$. This collective contribution is mostly unrelated to the critical mode responsible for stripe ordering (wavevector $q\equiv M$, which instead softens only at the MIT, see inset and Fig. \ref{fig:boson}b).
It arises instead from a diffuse region near the edges of the Brillouin zone \cite{Mahmoudian15},
indicative of the existence of many competing
 orders being frustrated by the long-range interactions \cite{Schmalian00,Mahmoudian15}. Translated to real space, these zone-boundary features correspond to the local (short-distance) dipoles postulated in Ref. \cite{Emery93}, arising from frustrated charge correlations. 
The fact that within the pseudogap phase the metallic character measured by the Drude weight, $D/D_0$, is larger than that implied by the one-particle residue $Z$ alone, as observed in Fig. \ref{fig:pseudogap}c, suggests that such collective modes could be actively contributing to charge transport as an additional conduction channel \cite{Fratini09,Caprara02}.

\paragraph{Concluding remarks.}

The existence of self-generated randomness with a suppressed energy scale  implies an equally suppressed temperature scale at which quantum coherence is lost. 
The collapse onto 
classical behavior should be further enhanced by the fact that the random potentials possess a continuous spectrum (Fig. \ref{fig:pseudogap}b), thus providing a natural source of electron decoherence \cite{CaldeiraLeggett}.
This could explain, for example, the puzzling behavior observed in the quarter-filled organic compounds $\theta$-(BEDT-TTF)$_2$X.  In these materials, the electron liquid shows precursors of glassiness despite the absence of structural disorder  \cite{Nad07,KagawaNphys13}, 
that are surprisingly well-captured by classical models \cite{Mahmoudian15}. 
Moreover, in agreement with the results found here, these materials display frustrated metastable orders
(seen as diffuse spots in X-ray diffraction images) 
that compete with the 
stripes \cite{KagawaNphys13}. Above an extremely low Fermi temperature, $T^*_{FL}\sim 20$K, which is two orders of magnitude lower than predicted by band-structure arguments,  the resistivity displays strange metal behavior with an approximately linear temperature dependence \cite{Takenaka05,SatoNMat19} compatible with strong scattering by low-energy bosonic modes.  The system also features a displaced Drude peak in the optical conductivity, suggestive of disorder-induced localization,
indicating that self-generated randomness could also be playing  a key role in the charge transport mechanism \cite{Kaveh82,Fratini16}.

Due to the general nature of the effects 
revealed here, it will be interesting to investigate their relevance in other quantum materials exhibiting bad metallic behavior \cite{Bruin13,Delacretaz17}, including those near integer fillings where long-range interactions are customarily neglected. In these systems, the reduced screening ability of electrons at the onset of the Mott transition should imply that long-range potentials play a significant role \cite{Emery93,Schmalian00,Padhi18},
therefore contributing to their 
anomalous thermodynamic and transport properties: 
the importance of long-range interactions and the ensuing nearly-classical behavior of charge fluctuations could bring the $T$-linear behavior of the resistivity, which characterizes correlated electrons at very high temperatures \cite{Mousatov19}, down to the experimentally relevant temperature range. Generally speaking, the interplay of Wigner and Mott physics should provide a promising new direction in research on strongly correlated materials.

\medskip 
\textit{Acknowledgments}\\
We thank S. Ciuchi, L. de' Medici and V. Dobrosavljevi\'c for enlightening discussions.
K. D. acknowledges the European Union's Horizon 2020 research and innovation program under the Marie Sk{\l}odowska-Curie grant agreement No. 754303.

$^*$ simone.fratini@neel.cnrs.fr

\pagebreak

\appendix

\setcounter{figure}{0}
\setcounter{equation}{0}
\renewcommand{\theequation}{S\arabic{equation}}
\renewcommand{\thefigure}{S\arabic{figure}}

\section*{Supplementary Information}

\section{Exact diagonalization}
\begin{figure*}
    \centering
    \newcommand*\rows{13}
    \subfigure[]
    {
    \resizebox{\columnwidth}{0.578\columnwidth}{
    \begin{tikzpicture}[scale=0.45]
    
    \foreach \row in {0, 1, ...,\rows} {
        \draw [gray,ultra thin] ($\row*(0.5, {0.5*sqrt(3)})$) -- ($(\rows,0)+\row*(0.5, {0.5*sqrt(3)})$);
        \draw [gray,ultra thin] ($\row*(1, 0)$) -- ($(\rows/2,{\rows/2*sqrt(3)})+\row*(1.0,{0*sqrt(3)})$);
        \draw [gray,ultra thin] ($\row*(1, 0)$) -- ($(0,0)+\row*(0.5,{0.5*sqrt(3)})$);
    }
    \foreach \row in {0,1, ...,\rows} {
        \draw [gray,ultra thin] ($\rows*(0.5,{0.5*sqrt(3)}) + \row*(1,0)$) -- ($\rows*(1,0)+\row*(0.5,0.5*sqrt(3)$) ;
    }
    
    \draw [black,ultra thick] ($(3.0,{1.0*sqrt(3)})$) -- ($(5.0,{1.0*sqrt(3)})$) ;
    \draw [black,ultra thick] ($(3.0,{1.0*sqrt(3)})$) -- ($(2.5,{1.5*sqrt(3)})$) ;
    \draw [black,ultra thick] ($(2.5,{1.5*sqrt(3)})$) -- ($(3.5,{2.5*sqrt(3)})$) ;
    \draw [black,ultra thick] ($(3.5,{2.5*sqrt(3)})$) -- ($(4.5,{2.5*sqrt(3)})$) ;
    \draw [black,ultra thick] ($(4.5,{2.5*sqrt(3)})$) -- ($(5.5,{1.5*sqrt(3)})$) ;
    \draw [black,ultra thick] ($(5.5,{1.5*sqrt(3)})$) -- ($(5.0,{1.0*sqrt(3)})$) ;
    \node[draw] at ($(3.25,{0.5*sqrt(3)})$) [fill=white,rounded corners=0.1cm] {$N_s=12$} ;
    
    
    \draw [black,ultra thick] ($(5.5,{3.5*sqrt(3)})$) -- ($(10.5,{3.5*sqrt(3)})$) ;
    \draw [black,ultra thick] ($(5.5,{3.5*sqrt(3)})$) -- ($(6.5,{4.5*sqrt(3)})$) ;
    \draw [black,ultra thick] ($(6.5,{4.5*sqrt(3)})$) -- ($(11.5,{4.5*sqrt(3)})$) ;
    \draw [black,ultra thick] ($(10.5,{3.5*sqrt(3)})$) -- ($(11.5,{4.5*sqrt(3)})$) ;
    \node[draw] at ($(8.5,{4.0*sqrt(3)})$) [fill=white,rounded corners=0.1cm] {$N_s=18$} ;
    
    \draw [black, ultra thick] ($(6.5,0.5*sqrt(3)$) -- ($(8.5,0.5*sqrt(3)$) ; 
    \draw [black, ultra thick] ($(6.5,0.5*sqrt(3)$) -- ($(6.0,{1.0*sqrt(3)})$) ;
    \draw [black, ultra thick] ($(6.0,{1.0*sqrt(3)})$) -- ($(7.0,{2.0*sqrt(3)})$) ;
    \draw [black, ultra thick] ($(7.0,{2.0*sqrt(3)})$) -- ($(9.0,{2.0*sqrt(3)})$) ;
    \draw [black, ultra thick] ($(9.0,{2.0*sqrt(3)})$) -- ($(10.0,{3.0*sqrt(3)})$) ;
    \draw [black, ultra thick] ($(10.0,{3.0*sqrt(3)})$) -- ($(11.0,{3.0*sqrt(3)})$) ;
    \draw [black, ultra thick] ($(11.0,{3.0*sqrt(3)})$) -- ($(12.0,{2.0*sqrt(3)})$) ;
    \draw [black, ultra thick] ($(12.0,{2.0*sqrt(3)})$) -- ($(11.5,{1.5*sqrt(3)})$) ;
    \draw [black, ultra thick] ($(11.5,{1.5*sqrt(3)})$) -- ($(9.5,{1.5*sqrt(3)})$) ;
    \draw [black, ultra thick] ($(9.5,{1.5*sqrt(3)})$) -- ($(8.5,0.5*sqrt(3)$) ;
    \node[draw] at ($(11.25,{0.75*sqrt(3)})$) [fill=white,rounded corners=0.1cm] {$N_s=24$} ;
    
    \draw [black,ultra thick] ($(7.0,{5*sqrt(3)})$) -- ($(12.0,{5*sqrt(3)})$) ;
    \draw [black,ultra thick] ($(7.0,{5.0*sqrt(3)})$) -- ($(8.5,{6.5*sqrt(3)})$) ;
    \draw [black,ultra thick] ($(8.5,{6.5*sqrt(3)})$) -- ($(13.5,{6.5*sqrt(3)})$) ;
    \draw [black,ultra thick] ($(12.0,{5*sqrt(3)})$) -- ($(13.5,{6.5*sqrt(3)})$) ;
    \node[draw] at ($(10.0,{5.75*sqrt(3)})$) [fill=white,rounded corners=0.1cm] {$N_s=24$} ;
    
    \draw [blue,ultra thick,-{Straight Barb}] ($(13.0,{3.0*sqrt(3)})$) -- ($(16.0,{3.0*sqrt(3)})$) ;
    \draw [blue,ultra thick,-{Straight Barb}] ($(13.0,{3.0*sqrt(3)})$) -- ($(14.5,{4.5*sqrt(3)})$) ;
    \node[draw] at ($(14.0,{2.5*sqrt(3)})$) [fill=white,rounded corners=0.1cm] {$t_c$} ;
    \node[draw] at ($(15.0,{5.0*sqrt(3)})$) [fill=white,rounded corners=0.1cm] {$t_p$} ;
    
    \end{tikzpicture}
    }
    \vspace{10\baselineskip}
    \label{fig:triangularlatticereal}
    }
    \subfigure[]
    {
    \includegraphics[width=6cm]{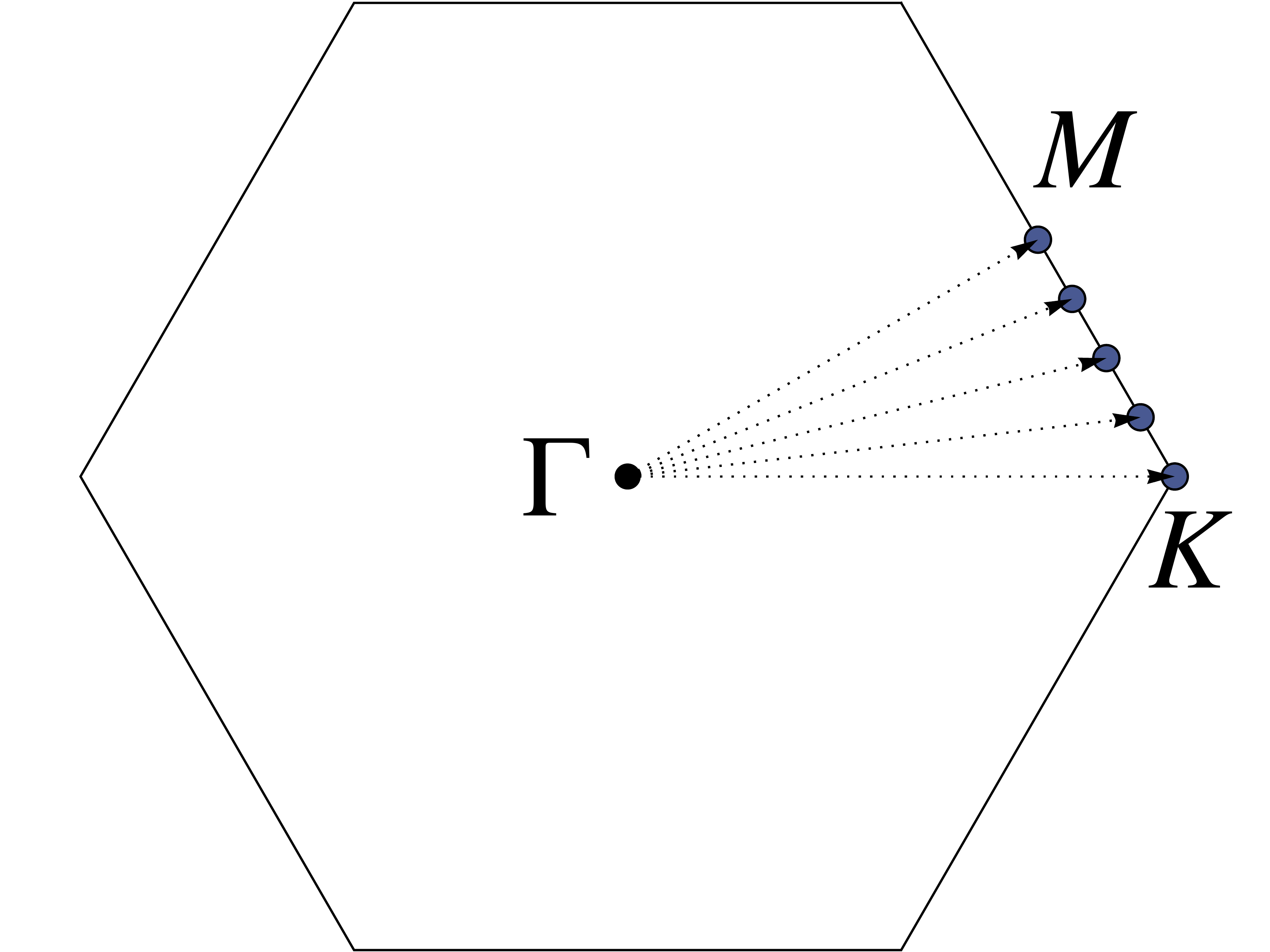}
    \vspace{10\baselineskip}
    \label{fig:triangularlatticeBZ}
    }
    \caption{(a) Sketch of cluster definitions for $N_s=12,18\:\mathrm{and}\:24$ sites. (b) Brillouin zone corresponding to the triangular lattice. The dotted lines are the paths along which the occupation number $\expval{\hat{n}_{\mathbf{k}}}$ is calculated for the determination of the quasiparticle renormalization $Z$.}
    \label{fig:triangularlattice}
\end{figure*}

All observables reported in this paper, unless explicitly noted otherwise, were computed via zero-temperature exact diagonalization calculations on an isotropic triangular lattice. When the Hilbert space of the system was sufficiently small, the Hamiltonian was diagonalized via brute force exact diagonalization. Otherwise, the Lanczos algorithm, complete with Gram-Schmidt orthogonalization, was employed. Calculations were performed on clusters of size $N_s=12,18$ and $24$ sites. The hopping and interaction terms along the different bond directions (shown in Fig.~\ref{fig:triangularlattice}a) were taken to be isotropic (ie, $t_c=t_p=t$ and $V_c=V_p=V$). 
The cluster geometries, shown in Fig.~\ref{fig:triangularlattice}a, were chosen such that the clusters were compatible with both stripe (M point) and three-fold (K point) charge order except for the rectangular prescription for the 24 site system, which misses the K point.  The translation vectors, ${\bm{T}}_1$ and ${\bm{T}}_2$, for each cluster are given as follows:
\begin{itemize}
    \item $N_s\!=\!12$: $\bm{T}_1\!=\!( L , L )$,  $\bm{T}_2\!=\!( -L , 2L )$ with $L=2$
    \item $N_s\!=\!18$: $\bm{T}_1\!=\!( L , 0 )$,  $\bm{T}_2\!=\!( 0 , L/2 )$ with $L=6$
    \item $N_s\!=\!24$: $\bm{T}_1\!=\!( L , 0 )$,  $\bm{T}_2\!=\!( 0 , 2L/3 )$ with $L=6$
    \item $N_s\!=\!24$: $\bm{T}_1\!=\!( L , L )$,  $\bm{T}_2\!=\!( -L , L/2 )$ with $L=4$.
\end{itemize}
We applied translation symmetries in a standard manner where necessary to reduce the computational cost and to minimize any spurious effects of degenerate ground states \cite{mambrini_etude_2002}. The use of these symmetries reduces the size of the Hilbert space in a given symmetry sector by a factor approximately equal to the number of sites, $\mathcal{H} \rightarrow \mathcal{H} / N_s$. 

\subsection{Twisted boundary conditions}
We employed twisted boundary conditions (TBCs) to improve discretization errors inherent in the kinetic portion of the Hamiltonian and to lift the degeneracy of the ground state. As described in Ref.~\cite{poilblanc_twisted_1991}, TBCs correspond to the insertion of a flux along the directions of the lattice torus which thereby modifies the hopping term of the Hamiltonian,
\begin{equation}
    H_{\mathcal{K}}^{\mathrm{TBC}} = -t \sum_{\langle ij \rangle} e^{i \bm{r}_{ij}\cdot \bm{\kappa}} c_i^\dagger c_j + h.c.,
\end{equation}
where $ \bm{r}_{ij}$ is the vector connecting two lattice sites and
\begin{equation}
    \bm{\kappa} = 2\pi \bigg( \frac{\varphi_x}{L_x} \hat{\bm{x}} + \frac{\varphi_y}{L_y} \hat{\bm{y}} \bigg).
\end{equation}
We can define the phase acquired along a given direction by an angle $\theta_i$ that is written in terms of a vector potential,
\begin{equation}
    \theta_{i} = 2\pi \xi_{i} = \frac{2 \pi}{\varphi_0} \oint \Big( A_1 \Vec{\alpha}_1 + A_2 \Vec{\alpha}_2 \Big ) \cdot d\Vec{l},
\end{equation}
where $\Vec{\alpha}_i$ corresponds to the unit lattice translation vector along the $i$-th direction of the torus. The prefactor, $2\pi/\varphi_0$ corresponds to a constant factor $2\pi q/h c$, where we set $\varphi_0 = hc/q = 1$. 

Under periodic boundary conditions (PBCs, $\xi_1=\xi_2=0$), the system suffers from ground state degeneracy issues which is well-known to lead to instabilities when treated via Lanczos diagonalization. Therefore, we introduced a small shift $\Vec{\xi} = (1.07654\times 10^{-4},-1.98673\times 10^{-4})$ to lift the degeneracy in the vicinity of the PBCs (and other highly degenerate boundary condition points) such that diagonalization via the Lanczos algorithm could be employed. TBCs not only remedy issues caused by degeneracy in the system, but also allow us to compute observables more accurately over a grid of flux points, thereby ensuring that our results approach the thermodynamic limit \cite{gros_boundary_1992,poilblanc_twisted_1991}. 

\subsection{Ewald summation}
For the interacting portion of the Hamiltonian [Eq. (1) of the main text], the Ewald summation was utilized to accurately implement a long-ranged potential on a discrete lattice of the form,
\begin{equation}
    V(\mathbf{r}_{ij}) = \sum_{\bm{n}} \frac{1}{\vert{\mathbf{r}_{ij}+\bm{n}}\vert^{\alpha}}, \label{eq:Ewaldpot}
\end{equation}
where $\bm{n} = n_1 \bm{T}_1 + n_2 \bm{T}_2$ with $n_i \in \mathbb{N}$. 
As in Ref.~\cite{pramudya_nearly_2011}, we make use of the integral representation of such a potential,
\begin{equation}
    \frac{1}{\vert r \vert^{\alpha}} = \frac{1}{\Gamma(\alpha/2)} \bigg( \int_0^{\varepsilon} dt \: t^{\frac{\alpha}{2}-1} e^{-r^2t} + \int_{\varepsilon}^\infty dt \: t^{\frac{\alpha}{2}-1} e^{-r^2t} \bigg),
\end{equation}
to obtain an expression of the form
\begin{multline}
    V(\bm{r}) = \frac{\pi^{d/2}c^{(\alpha-d)/2}}{\mathrm{vol}} \frac{1}{\Gamma(\alpha/2)} \sum_{\bm{k}}\Bigg[ \Big( \cos(\bm{k}\cdot \bm{r}) - 1 \Big) \\
    \times \phi_{\frac{d-\alpha}{2}-1}\bigg( \frac{\vert{k}\vert^2}{4c} \bigg)\Bigg] \\
    + \frac{c^{\alpha/2}}{\Gamma(\alpha/2)} \sum_{\bm{n}} \Bigg[ \phi_{\frac{\alpha}{2}-1} \Big( c \vert \bm{r}+\bm{n} \vert^2 \Big) \\
    - \phi_{\frac{\alpha}{2} - 1}\Big( c \vert \bm{r}\vert^2  \Big) \Bigg] 
    + \frac{c^{\alpha/2}}{\Gamma(\alpha/2)}\frac{2}{\alpha} - \frac{1}{\vert \bm{r} \vert^\alpha},
\end{multline}
where 
\[
  \phi_{p}(a) =
  \begin{cases}
                                   0 & \text{if $a=0$} \\
                                   \int_1^{\infty} dt \: t^p e^{-at} & \text{if $a\neq0$}. \\
  \end{cases}
\]
This general expression is valid in any dimension $d$ and for any range of interaction, $\alpha$. Each of the summation terms in this expression converges rapidly in reciprocal and real space, respectively. All calculations were performed at convergence of the potential.

\section{Finite size convergence}

\subsection{Drude weight}
\begin{figure}[t]
    \centering
    \includegraphics[width=5.5cm]{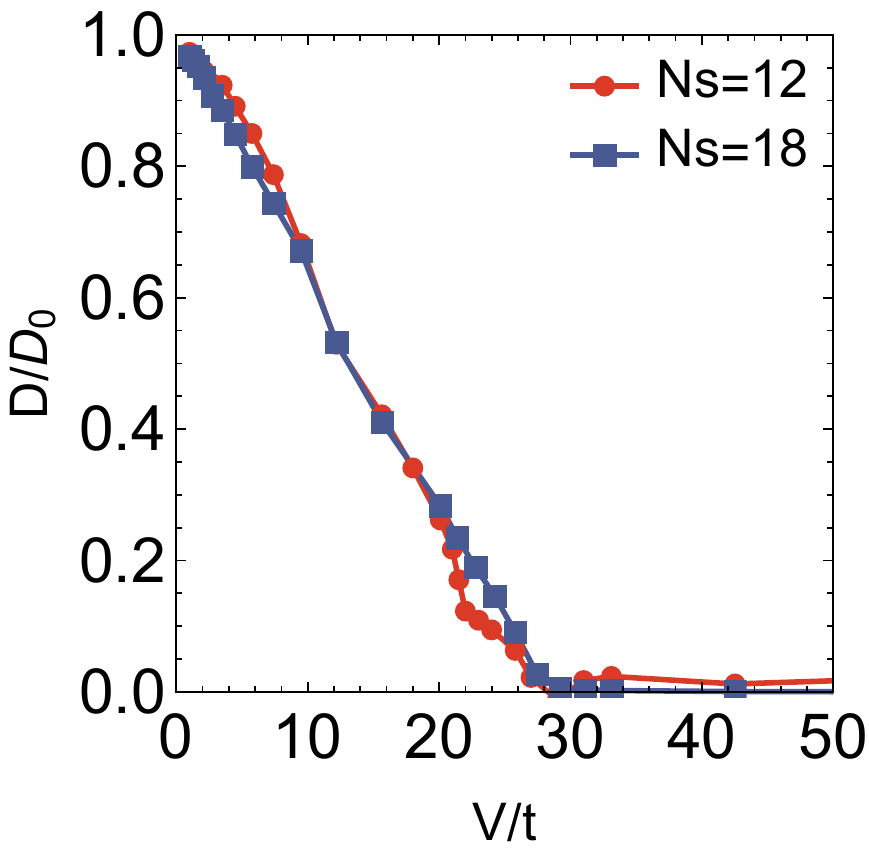}
    \caption{The Drude weight, given as a fraction of the non-interacting value, for the $N_s=12$ and $18$ site systems with unscreened Coulomb interactions $V(R)\propto 1/R^\alpha$ ($\alpha=1$). Both clusters display a transition from a metallic to insulating state at $(V/t)_c \approx 29$. }
    \label{fig:drudefinite}
\end{figure}

\begin{figure}[t]
    \centering
    \includegraphics[width=\columnwidth]{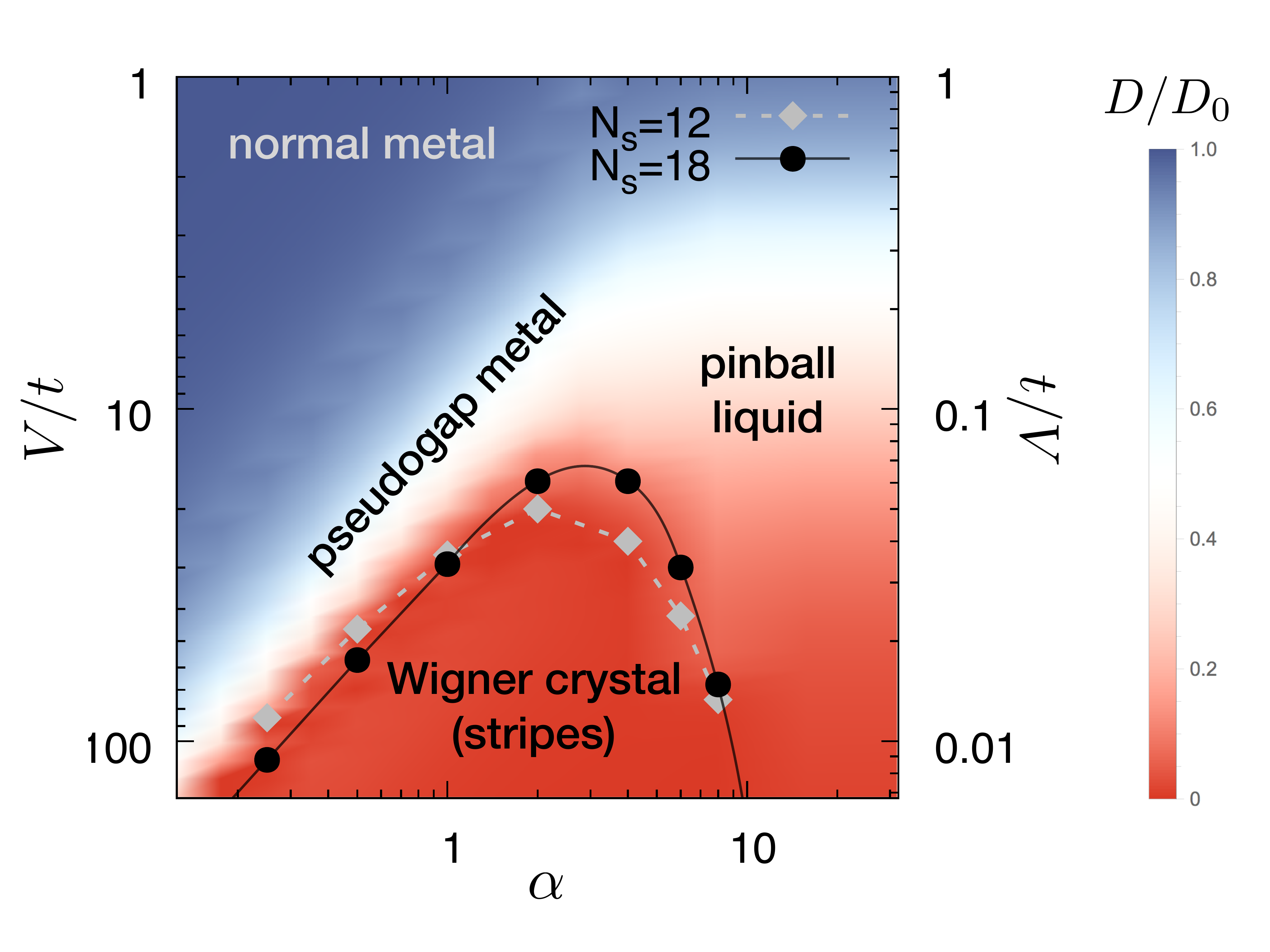}
    \caption{Phase diagram showing metal-insulator transition points determined as the disappearance of the Drude weight ($D/D_0$) on the $N_s=12$ and $18$ site systems (gray and black points respectively). Both systems possess similarly dome-shaped structures of the stripe-ordered phase, showing that the agreement between different system sizes is consistent over multiple ranges of interaction, $\alpha$.}
    \label{fig:phasediagramfinite}
\end{figure}

The Drude weight is calculated as in Ref.~\cite{koretsune_exact_2007} as
\begin{equation}
    \frac{D_{\mu}}{2\pi e^2} = \frac{1}{2N}\matrixelement{0,\varphi}{F_{\mu \mu}}{0,\varphi} + \frac{1}{N}\sum_{n \neq 0} \frac{\abs{\matrixelement{n,\varphi}{J_{\mu}}{0,\varphi}}^2}{E_0(\varphi)-E_n(\varphi)}
\end{equation}
where 
\begin{equation}
    J_{\mu} = \frac{\partial H}{\partial \varphi_{\mu}} \quad\mathrm{and}\quad F_{\mu \nu} = \frac{\partial^2 H}{\partial \varphi_{\mu} \partial \varphi_{\nu}}
\end{equation}
and $\mu, \nu$ represent directions along the lattice. The notations $\ket{0,\varphi}$ and $\ket{n,\varphi}$ indicate the ground state and the $n$-th excited state, respectively, for a given flux, $\varphi=(\varphi_x,\varphi_y)$. The Drude weight reported in the text has been calculated in the direction $c$ of the rotationally symmetric $N_s=12$ lattice (see Fig.~\ref{fig:triangularlattice}a), and  upon averaging over the directions $\theta=0,\pi/6$ and $\pi/3$ for the $N_s=18$ lattice ($\theta$ being the angle with respect to c). 
As shown in Figs.~\ref{fig:drudefinite} and ~\ref{fig:phasediagramfinite}, the Drude weight (given as a fraction of the non-interacting value, $D/D_0$) agrees remarkably well for the $N_s=12$ and $18$ site clusters, suggesting that our calculations are well converged to the thermodynamic limit. For each of the 697 points in $(\alpha,t/V)$ used to construct the colormap reported in Fig.~\ref{fig:phasediagramfinite} and in Fig. 1 of the main text, we averaged $D/D_0$ over a 20 x 20 grid of flux points for the $N_s=12$ site system.

The values of $D/D_0$ reported in Figs. 1 and 2 of the main text for $N_s=18$ were calculated via Lanczos diagonalization with the use of translation symmetries to determine the ground state symmetry sector. An 11 x 11 grid of flux points was used, with small shifts introduced (as discussed previously) to minimize degeneracy effects. Despite the use of symmetries and shifted flux points, some spurious degeneracy effects were observed that led to the removal of a subset of points from the average. This removal was conducted by implementing a cutoff and discarding unphysical values below the cutoff ($D/D_0 < -0.05$). For any given $\alpha$ and $V/t$, at most 30 points were discarded which still yields an accurate averaging over 91 flux points.

\subsection{Spectral function}\label{sec:spectral-function}
The spectral function, reported from calculations on the $N_s=18$ site system, was calculated as
\begin{multline}
    A(\omega) = -\frac{1}{\pi} \mathrm {Im} \ \sum_{i} \sum_{n \neq 0} \frac{\vert \matrixelement{\Psi_n^{N+1}}{c_{i}^\dagger}{\Psi_0^N} \vert^2}{\omega - (E_n^{N+1} - E_0^N) + i 0^+} \\+ \frac{\vert \matrixelement{\Psi_n^{N-1}}{c_{i}}{\Psi_0^N} \vert^2}{\omega - (E_0^N - E_n^{N-1}) + i 0^+},
\end{multline}
where the summation over $i$ indicates a sum over the discrete lattice sites in real space, $\mathbf{r}_i$. The subscripts on the wave functions indicate the ground or $n$-th excited state, while the superscripts indicate the number of particles. For each value of $V/t$, the single-particle spectral function was averaged over a 4 x 4 grid of flux points and a Gaussian broadening was applied. The spectral function was already well-converged for the 4 x 4 grid of flux points and increasing the size of the mesh to 11 x 11 did not qualitatively change the results.

\begin{figure}[t]
    \centering
    \includegraphics[width=6.5cm]{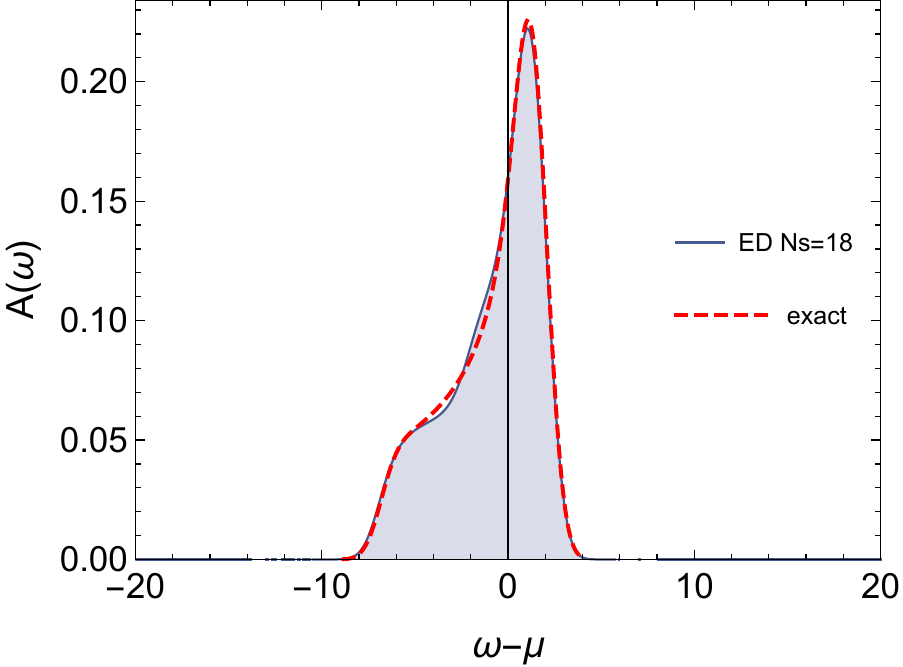}
    \caption{A comparison of the non-interacting spectral function on the triangular lattice in the thermodynamic limit (red, dashed line) and averaged over 121 twisted boundary conditions for a calculation on the $N_s=18$ site cluster (blue). The same Gaussian broadening $\delta=0.7t$ has been applied in both cases.}
    \label{fig:AwV0}
\end{figure}
Fig. \ref{fig:AwV0} shows our ED result for the spectral function in the non-interacting limit obtained on the 18-site triangular lattice with an 11x11 grid of flux points (blue), recovering very accurately  the known spectral function in the thermodynamic limit (red dashed). 

\subsection{On-site potential distribution}
\begin{figure}[t]
    \centering
    \includegraphics[width=5.5cm]{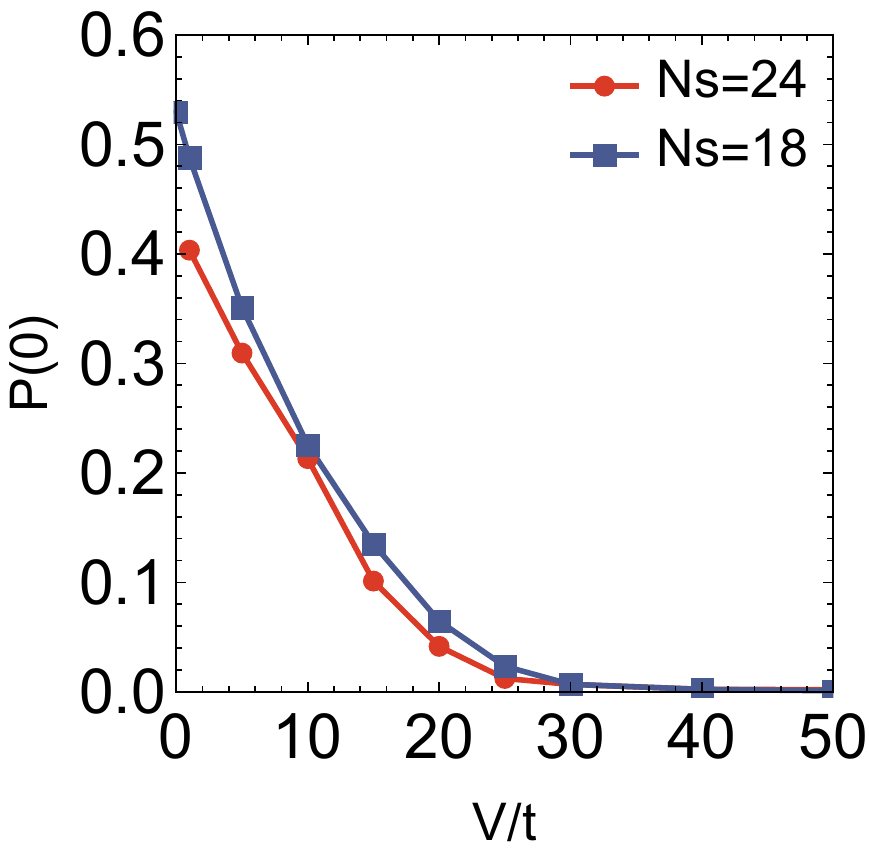}
    \caption{On-site potential distribution $P(\phi)$ calculated at  $\phi=0$, illustrating the progressive opening of the Coulomb gap. The results  for $N_s\!=\!18$ and $24$ sites are in very good quantitative agreement.}
    \label{fig:Pof0finite}
\end{figure}

The distribution of on-site potentials, $P(\phi_i)$, defined as 
\begin{equation}
    P(\phi_i) = \matrixelement**{\psi}{\delta\Bigg( \phi_i - \sum_{j\neq i} V(R_{ij})\big(\hat{n}_j-\langle \hat{n} \rangle\big) \Bigg)}{\psi},
\end{equation}
was computed on a 4 x 4 grid of flux points for the $N_s\!=\!18$ and $24$ site systems. Histograms were constructed from the data on the flux grid with a width $\phi/V\!=\!0.002$, and a Gaussian broadening $\phi/V\!=\!0.05$ was applied to obtain smooth curves. Fig.~\ref{fig:Pof0finite} illustrates the central value $P(0)$ vs $V/t$, showing very good agreement between $N_s\!=\!18$ and $24$ sites.

\subsection{Quasiparticle weight}
The quasiparticle weight, $Z$, is calculated as the size of the discontinuity of the occupation function, $\expval{n_{\mathbf{k}}}$. TBCs allow us to compute $\expval{n_{\mathbf{k}}}$ along a finely discretized path in momentum space. For each path, we determine $\expval{n_{\mathbf{k}}}$ at $\Gamma$ and apply a flux such that $\mathbf{k}$ would lie along the line in momentum space dictated by the path. The results showed some degree of anisotropy, stemming from an anisotropic quasiparticle renormalization, and as a result, we present $Z$ computed as an angular average over five different paths, shown in Fig.~\ref{fig:triangularlattice}b. 

\subsection{Buildup of short-range correlations}
\begin{figure}[ht]
    \centering
    \includegraphics[width=6cm]{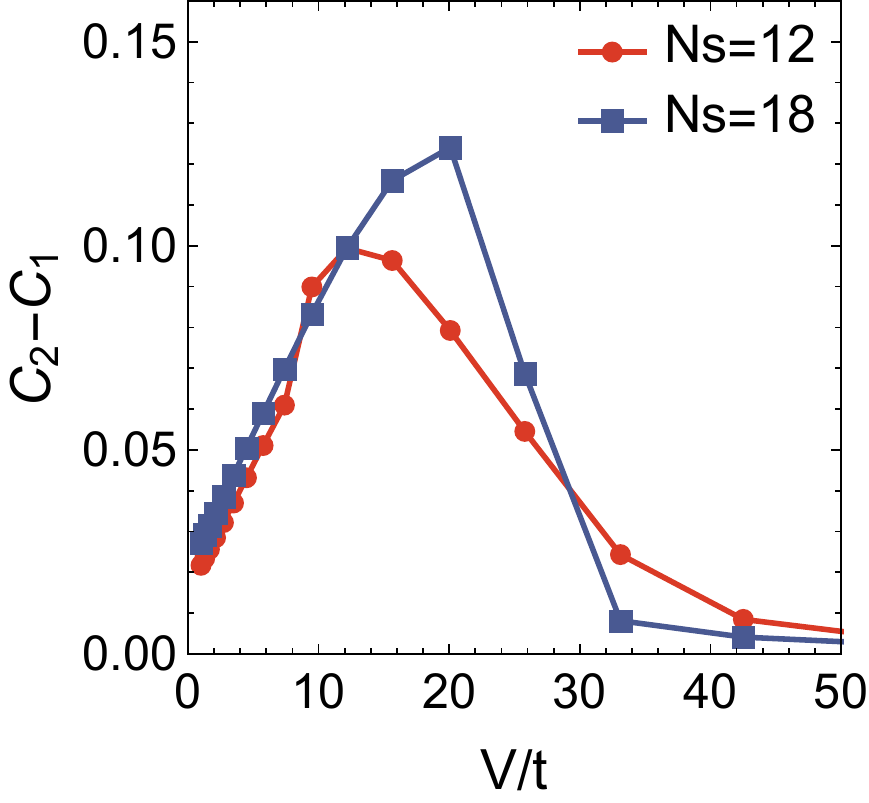}
    \caption{Difference of the charge-charge correlation function $C(i,j)$  computed on the second and on the first shell of neighbors, illustrating the buildup of the correlation hole.}
    \label{fig:C2C1finite}
\end{figure}

To study the behavior of short-range correlations, we  compute the charge correlation function in real space,
\begin{equation}
    C(i,j) = \sum_{i,j}\matrixel{\psi_0}{\hat{n}_i \hat{n}_j}{\psi_0}.
\end{equation}
Fig.~\ref{fig:C2C1finite} reports the difference between $C(i,j)$ on the  second shell of neighbors (next nearest neighbors) and the first shell (nearest neighbors). A progressive depletion of the first shell and transfer to longer distances marks the buildup of short-range correlations, corresponding to the formation of a correlation hole around each charge carrier.
The effect steadily increases with $V/t$, paralleling the quasiparticle renormalization observed in Fig. 2c of the main text, until it is eventually interrupted by the establishment of long-range stripe correlations at the metal-insulator transition.

\section{Analytical estimates of the transition lines in Fig.1.}
Here we provide a description of the analytical estimates for the transition lines obtained for small and large $\alpha$, shown in the dashed and dotted lines, respectively in Fig. 1. 

At small $\alpha$, melting of the stripe phase occurs through proliferation of defects. The energy of a single defect, $E_d$, can be evaluated by computing the increase in electrostatic energy, as per Eq. (\ref{eq:Ewaldpot}), upon moving a single particle from an occupied to a neighboring empty site. 
The quantum melting transition  occurs when $t \sim E_{d} $. 
From  the asymptotic expression $E_{d}\simeq  0.469 V \alpha$  we obtain $(V/t)_c \propto 1/\alpha$ for small $\alpha$. This behavior is well reproduced by the ED results, as shown in Fig. 1 (black dashed line). 

At large $\alpha$, the stripes melt into a pinball liquid, which  possesses a marked short-range threefold ordering. In this case the transition point  can be estimated by comparing directly the energies of the two phases for large $\alpha$, to lowest order in the kinetic term $\propto t$. The Madelung energy of the pinball liquid and that of the stripes are equal for short-range interactions. However, as soon as $\alpha < \infty$, the Madelung energy of the pinball liquid becomes higher due to contributions from the interactions between next nearest neighbors and beyond.  The difference in potential energy between the pinball and stripe charge configurations behaves asymptotically as $\Delta E \propto V/(R_2)^\alpha $, with $R_2=\sqrt{3}$ the second neighbor distance. The gain in kinetic energy due to quantum fluctuations at finite $t$ is proportional to $t^2/V$ in the stripe phase,  while it is larger in the pinball liquid, being linear in $t$. Equating these two contributions  leads to $(V/t)_c\propto 3^{\alpha/2}$, represented as a dotted line in Fig. 1. 

\section{Pseudogap physics on the square lattice}
\begin{figure}[h]
    \centering
    \includegraphics[width=\columnwidth]{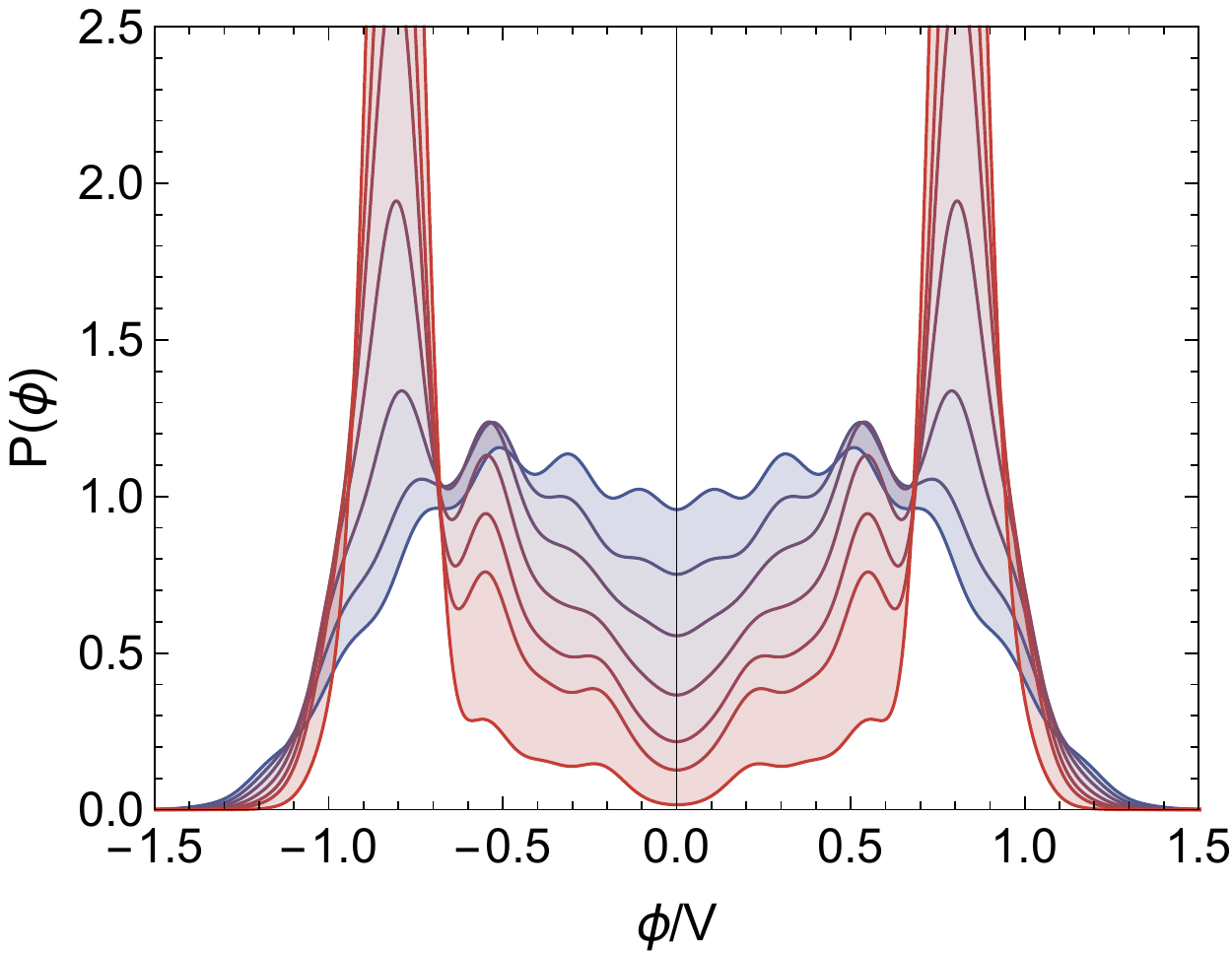}
    \caption{The development of the pseudogap in the spectral function on the square lattice for $\alpha=1$ as $V/t$ is increased (gradient of blue to red lines; the corresponding values of $V/t$ are indicated in the figure legend). The inset shows the value of $A(\omega)$ at $\omega=\mu$ which steadily decreases before a hard gap opens at $V/t \approx 8$. The number of sites, $N_s=18$, is the same as in Fig. 2 of the main text. These results are averaged over an 11 $\times$ 11 grid of flux points.}
    \label{fig:square-pseudogap}
\end{figure}

In order to demonstrate the generality of the behavior discovered, we present here the results of the spectral function on the square lattice. Although this lattice geometry does not have any inherent frustration, we still observe the development of a pseuodogap in the spectral function with increasing $V/t$, shown in Fig.~\ref{fig:square-pseudogap}. The spectral function was calculated as explained previously in Sec.~\ref{sec:spectral-function}. The translation vectors for the $N_s=18$ site cluster used are given as: 
\begin{itemize}
    \item $N_s\!=\!18$: $\bm{T}_1\!=\!( L , L )$,  $\bm{T}_2\!=\!( -L , L )$ with $L=3$.
\end{itemize}

We argue that long-range interactions are the source of frustration responsible for driving the development of the pseudogap in both the triangular and square lattices. The collective behavior induced by these interactions involves the coordination of many sites regardless of the local lattice geometry.  It is also interesting to note that Refs.~\cite{pramudya_nearly_2011,rademaker_suppressed_2018, fratini_unconventional_2009} have demonstrated frustration of charge order arising from long-range interactions  in otherwise non-frustrated lattices (square, cubic) in the classical limit. In summary, the development of strongly correlated behavior arising from long-range interactions is a general phenomenon and does not rely on the geometrical frustration of the lattice.  

\bibliography{SI-references.bib}

\end{document}